\documentclass[aip,amsmath,amssymb,preprint]{revtex4-1}

\usepackage{natbib}
\usepackage{graphicx}
\usepackage{graphics}
\usepackage{amsfonts}
\usepackage{amsmath}
\usepackage{amssymb}
\usepackage{color}
\usepackage{soul}
\usepackage{multirow}
\usepackage{chngcntr}
\usepackage{longtable} %https://www.overleaf.com/project/5c61386290b8dc3b1e3a4a6f
\usepackage{dcolumn}
\usepackage{bm}

\usepackage[utf8]{inputenc}
\usepackage[T1]{fontenc}
\usepackage{mathptmx}

\begin{document}

\preprint{AIP/123-QED}

\title{Efficient prediction of Nucleus Independent Chemical Shifts for polycyclic aromatic hydrocarbons}

\author{Dimitrios Kilymis}

\affiliation{\small CIRIMAT, Universit\'e de Toulouse, CNRS, Universit\'e Toulouse 3 - Paul Sabatier, 118 Route de Narbonne, 31062 Toulouse cedex 9, France}
\affiliation{\small R\'eseau sur le Stockage \'Electrochimique de l'\'Energie (RS2E), F\'ed\'eration de Recherche CNRS 3459, HUB de l'\'Energie, Rue Baudelocque,  80039 Amiens, France}

\author{Albert P. Bart\'{o}k}

\affiliation{\small Warwick Centre for Predictive Modelling, Department of Physics and School of Engineering, University of Warwick, Coventry, CV4 7AL, United Kingdom}
\affiliation{\small Rutherford Appleton Laboratory, Scientific Computing Department, Science and Technology Facilities Council, Didcot, OX11 0QX, United Kingdom}

\author{Chris J. Pickard}

\affiliation{\small Department of Materials Science and Metallurgy, University of Cambridge, UK}
\affiliation{\small Advanced Institute for Materials Research, Tohoku University, Aoba, Sendai 980-8577, Japan}

\author{Alexander C. Forse}

\affiliation{\small Department of Chemistry, University of Cambridge, Lensfield Road, Cambridge, CB2 1EW, UK}
\affiliation{\small Department of Chemistry, Department of Chemical and Biomolecular Engineering, and Berkeley Energy and Climate Institute, University of California, Berkeley, CA94720, USA}

\author{C\'eline Merlet}

\email{merlet@chimie.ups-tlse.fr}

\affiliation{\small CIRIMAT, Universit\'e de Toulouse, CNRS, Universit\'e Toulouse 3 - Paul Sabatier, 118 Route de Narbonne, 31062 Toulouse cedex 9, France}
\affiliation{\small R\'eseau sur le Stockage \'Electrochimique de l'\'Energie (RS2E), F\'ed\'eration de Recherche CNRS 3459, HUB de l'\'Energie, Rue Baudelocque,  80039 Amiens, France}

\begin{abstract}
Nuclear Magnetic Resonance (NMR) is one of the most powerful experimental techniques to characterize the structure of molecules and confined liquids. Nevertheless, the complexity of the systems under investigation usually requires complementary computational studies to interpret the NMR results. In this work we focus on polycyclic aromatic hydrocarbons (PAHs), an important class of organic molecules which have been commonly used as simple analogues for the spectroscopic properties of more complex systems, such as porous disordered carbons. We use Density Functional Theory (DFT) to calculate $^{13}$C chemical shifts and Nucleus Independent Chemical Shifts (NICS) for 34 PAHs. The results show a clear molecular size dependence of the two quantities, as well as the convergence of the $^{13}$C NMR shifts towards the values observed for graphene. We then present two computationally cheap models for the prediction of NICS in simple PAHs. We show that while a simple dipolar model fails to produce accurate values, a perturbative tight-binding approach can be successfully applied for the prediction of NICS in this series of molecules, including some non-planar ones containing 5- and 7-membered rings. This model, one to two orders of magnitudes faster than DFT calculations, is very promising and can be further refined in order to study more complex systems.
\end{abstract}

\maketitle

\section{Introduction}

Nuclear Magnetic Resonance (NMR) is a powerful tool to study structural and dynamical properties in a wide range of systems including energy storage~\citep{Blanc13,Griffin16,wang2011}, biological systems~\citep{shen2008,Cavanagh_book,Marion13} and glasses~\citep{Youngman18,stebbins1997}. Indeed, the ability to probe specific nuclei and to conduct non invasive \emph{in situ} experiments makes it a method of choice for many applications. However, in a number of experiments, and increasingly so due to the growing complexity of the systems studied, the interpretation of the results is far from being straightforward and calculations of chemical shifts through different methods are nowadays very common. While standard Density Functional Theory (DFT) calculations are becoming faster and more accurate, their use is still limited to relatively small systems (a few hundred atoms)~\citep{schleder2019} and therefore the development of computationally cheap models for the prediction of chemical shifts can be very valuable. Several computationally affordable approaches to calculate or predict NMR parameters for hydrocarbons have been proposed over the years, ranging from ring current models~\citep{haigh1979}, to early neural-networks~\citep{kvasnicka1991,meiler2000} and the most recent boom of sophisticated machine-learning algorithms~\citep{paruzzo2018,gerrard2019,jonas2019}. However, the performance of the latter models has yet to be proven in the case of amorphous materials, such as disordered porous carbons, since these systems require structures of a few thousand atoms in order to be accurately described and, most importantly, contain rather unusual atomic topologies. This means that the current machine-learned predictive models for carbon, trained on data for isolated molecules, are not guaranteed to give accurate results for extended solids. Nevertheless, we should point out that there have recently been encouraging results using pure machine-learning approaches for the prediction of NMR parameters in other types of disordered systems, such as oxide glasses~\citep{cuny2016,chaker2019}.

Disordered porous carbons are a class of materials used in many applications such as energy storage, gas storage, water desalination and catalysis~\citep{Zhao06}. In all these applications, an accurate characterization of the structure of the carbon and the fluid adsorbed inside the porosity is of primary importance to understand and improve the performance of the systems. In the past, NMR has been proposed as a method to determine the pore size distribution of porous carbons~\citep{Xing14}, the size of aromatic carbon domains~\citep{Forse15b} and to study the ion dynamics in such confined environments\citep{Griffin14,Forse17,Borchardt18}. 
The extraction of structural and dynamical information from NMR experiments for ions or molecules adsorbed inside porous carbons is largely based on the existence of specific chemical shifts arising from secondary magnetic shieldings due to the presence of ring currents. These shielding effects are also of interest for the study of adsorption in different systems, such as metal-organic frameworks (MOFs)~\citep{nandy2018,zhang2018} or zeolites~\citep{white1992}. Concerning disordered porous carbons, there have been a number of works where the ring currents and the associated chemical shifts, the Nucleus Independent Chemical Shifts (NICS), have been estimated using DFT calculations on small aromatic molecules~\citep{Xing14,forse2014,resing1987} known as Polycyclic Aromatic Hydrocarbons (PAHs). PAHs are interesting to study as simplified models for disordered carbons, which are more challenging to describe from a theoretical point of view. However, these molecules are also interesting in their own right since they are present in fossil fuels and are a byproduct of incomplete combustion for a number of materials~\citep{Weisshoff02}. This makes their detection and identification important, especially since they can be toxic or have carcinogenic, mutagenic or teratogenic properties~\citep{harvey1988, umbuzeiro2008}. 

In this work, we report on the calculation of $^{13}$C chemical shifts as well as NICS for a number of PAHs using DFT. We first describe molecular size effects on the $^{13}$C NMR and NICS before proposing two models to predict NICS in a more efficient way. The first model is simply based on a classical dipolar model for the NICS~\citep{carrington1967} where the contributions of the different rings of the molecule to the chemical shielding are considered additive. We show that this model is insufficient as it does not take into account the change in local geometry and molecular susceptibility, and as it is valid only for relatively large probe distances. The second model is a perturbative tight-binding approach. We show that this model is much more accurate for a large range of molecules and discuss possible strategies to increase its accuracy for more complex cases.

\section{Models and methods}

\subsection{DFT calculations using Gaussian}

\begin{figure}[ht!]
\centering
\includegraphics[scale=1.0]{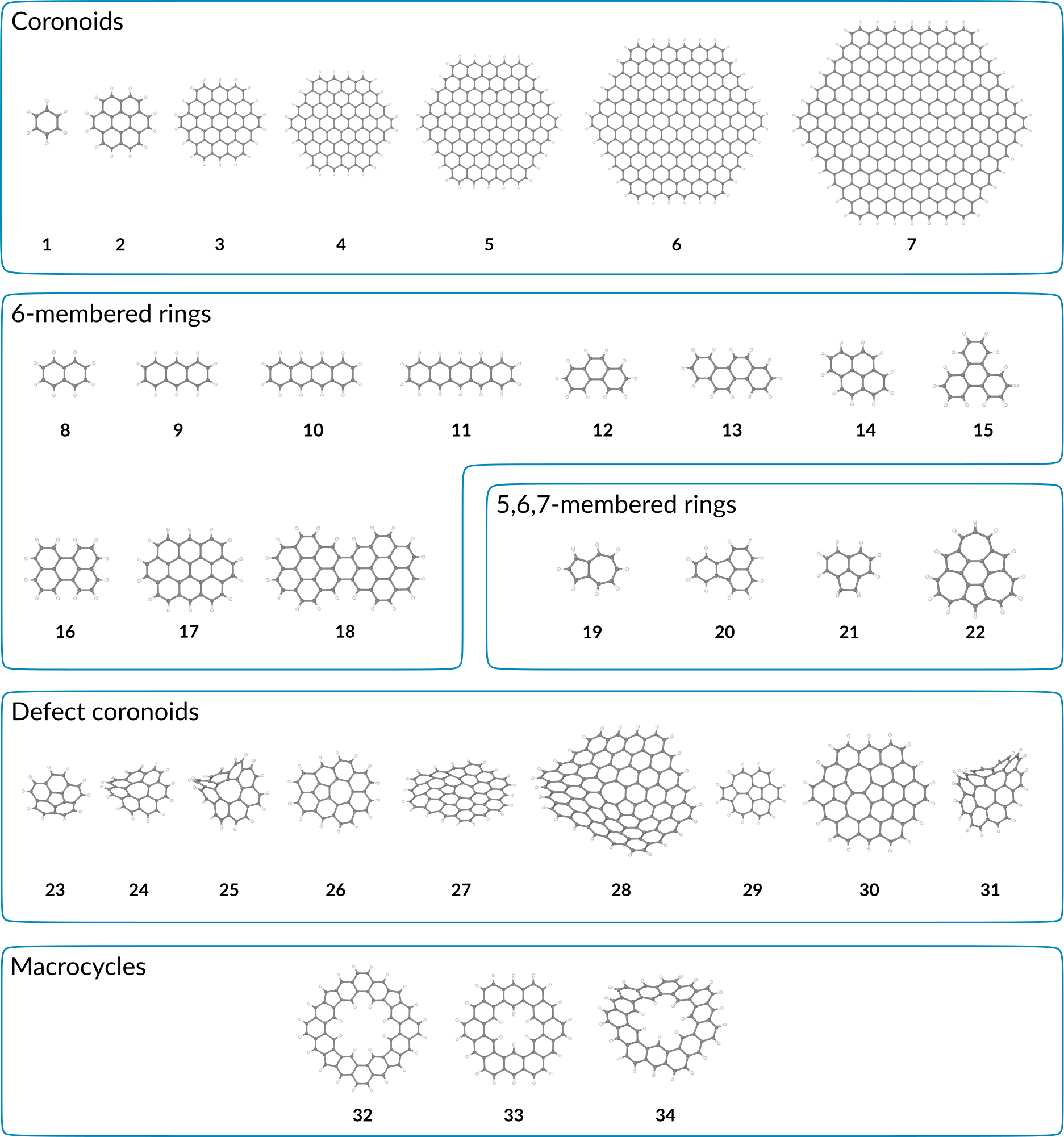}
\caption{All PAHs studied in this work.}
\label{fig:coronenes}
\end{figure}

We have carried out DFT calculations on a series of simple aromatic hydrocarbons using Gaussian 09~\citep{frisch2013}. Following earlier works~\citep{Moran03,forse2014}, we have used a 6-31G(d) basis set and the B3LYP exchange-correlation functional~\citep{becke1993}. A comparison of the performance of different basis sets shows, as reported previously~\citep{Moran03,Facelli06}, that the basis set has little influence on the NMR results (see Supplementary Material). After optimizing the structures at the lowest possible spin multiplicity, we calculated their vibrational frequencies in order to ensure that we have reached true minima, and proceeded to the calculation of the  $^{13}$C NMR shieldings. The chemical shifts calculated using DFT are given with respect to the value of tetramethylsilane (TMS) following:
\begin{equation}
    \delta^{13}{\rm C} = -(\sigma_{\rm C}-\sigma_{\rm TMS}) = -(\sigma_{\rm C}-\sigma_{\rm calc}^{\rm benzene}-\delta_{\rm TMS}^{\rm benzene})
\end{equation}
where $\sigma_{\rm C}$ is the calculated isotropic shielding of each carbon atom, $\sigma_{\rm calc}^{\rm benzene}$ the calculated isotropic shielding of the carbon atoms in benzene and $\delta_{\rm TMS}^{\rm benzene}$ is the experimental value for the chemical shift of benzene in TMS~\citep{gupta2005}. The isotropic shieldings have been calculated by averaging over the diagonal components of the shielding tensor. At the same time, we calculated the values of the isotropic NICS on external grid points (see Fig.~S2 for an example of grid). These are given by:
\begin{equation}
    {\rm NICS_{DFT}} = -\frac{1}{3} \left (\sigma_{xx,\mathrm{DFT}} + \sigma_{yy,\mathrm{DFT}} + \sigma_{zz,\mathrm{DFT}} \right )    
\end{equation}
where $\sigma_{\rm \alpha\alpha}$ are the diagonal elements of the NICS tensor. The grid points where the NICS values have been evaluated span from the molecule center up to a maximum distance of 15~\r{A} from the atom that is farthest from the center with a minimum 2~\r{A} spacing between the points. The values of the isotropic NICS calculated by DFT are designated as NICS$_{\rm DFT}$ in the remainder of the article. In total, we have studied 34 molecules which we divide in five groups depending on the type of rings present (5,6,7-membered rings) and the planarity of the molecule (see Fig.~\ref{fig:coronenes}). 

\subsection{The dipolar model}

Classical models of ring currents predict that the contribution of a single aromatic ring to the chemical shielding at a given point in space follows a dipolar dependence~\citep{carrington1967} in which the various components of the chemical shielding tensor are given by:
\begin{eqnarray}
    \sigma_{xx} &=& -\frac{1}{4 \pi}\chi_\perp(3\sin^2{\theta}-1)\frac{1}{r^3} \\
    \sigma_{yy} &=& \frac{1}{4 \pi}\chi_\perp\frac{1}{r^3} \\
    \sigma_{zz} &=& -\frac{1}{4 \pi}\chi_\parallel(3\cos^2{\theta}-1)\frac{1}{r^3}
\end{eqnarray}
where $\chi_\perp$ and $\chi_\parallel$ are the molar magnetic susceptibilities of the material in the directions perpendicular and parallel to the z-axis respectively, $r$ is the distance between the considered point and the centre of mass of a ring and $\theta$ is the angle between the vector that connects the point to the centre of mass and the vector normal to the ring centre (see Supplementary Material for a scheme showing these quantities). For $\chi_\perp$ and $\chi_\parallel$, we have used the values published by Ganguli and Krishnan for graphite~\citep{ganguli1941}, equal to –7.5~10$^{-11}$~m$^3$~mol$^{-1}$ and –3.2~10$^{-9}$~m$^3$~mol$^{-1}$ respectively. The NICS, which can be compared to DFT calculations is then given by:
\begin{equation}
    {\rm NICS_{dip}} = -\frac{1}{3} \left (\sigma_{xx,\mathrm{dip}} + \sigma_{yy,\mathrm{dip}} + \sigma_{zz,\mathrm{dip}} \right ).    
\end{equation}
The contributions for the different rings in a given molecule are simply added together. For example, for coronene, contributions from 7 rings are summed to estimate the total NICS.

\subsection{The tight-binding model}

The perturbative tight-binding (TB) model used to calculate the NICS is the one proposed by McWeeny~\citep{mcweeny1958}, which builds on the London theory for ring currents~\citep{london1937}. Whereas this perturbative approach has been used in the past for the calculation of chemical shifts in hydrocarbons~\citep{lazzeretti2000}, it has not yet been applied for the prediction of NICS. In this context, we consider a non-orthogonal tight binding (TB) Hamiltonian and compute the orbital energies from the generalised eigenvalue problem
\begin{equation}
\mathcal{H} \mathbf{c} = \varepsilon \mathbf{S} \mathbf{c}.
\end{equation}
The total energy of the system is obtained as the sum of the occupied orbital energies:
\begin{equation}
    E = \sum_i^\textrm{occ} \varepsilon_i.
\end{equation}
To study the magnetic response of planar aromatic hydrocarbons, which is the main focus of this work, we use the reduced Hamiltonian, $\mathcal{H}$, and overlap matrices, $\mathbf{S}$. The Hamiltonian of such systems in a minimal atomic basis is a block-diagonal matrix, and the aromatic electronic structure is represented by the block corresponding to the basis set of $p$ orbitals that are perpendicular to the mirror plane of the molecule, located on the carbon atoms. We expect that the magnetic response far from the molecule will be determined by the response of the aromatic subsystem, hence we restrict our model to this block of the Hamiltonian. We use the self consistent Fock and overlap matrices obtained by running B3LYP/STO-3G calculations on benzene, naphthalene and coronene molecules to fit the elements of the Hamiltonian and overlap matrices in the form of
\begin{equation}
\{\mathcal{H}_{ij},S_{ij}\} = \begin{cases}
A \exp( B r_{ij}^2 ) &\textrm{ \quad for \quad} i\ne j\\
C(n_i)  &\textrm{ \quad for \quad} i = j\\
\end{cases}
\label{eq:TB}
\end{equation}
where $n_i$ is the number of nearest carbon neighbours of carbon atom $i$.
Values of the parameters in equation~\ref{eq:TB} are given in Table~\ref{tab:TB}.

\begin{table} [ht!]
{\setlength{\tabcolsep}{0.21cm}
\begin{tabular}{l c c }
\hline
       & $\mathcal{H}$ & $S$ \\
\hline
$A$    & $-0.38840357$   & $0.61394025$ \\
$B$    & $-0.47459851$   & $-0.54122058$ \\
$C(2)$ & $-0.107001$     & $1$ \\
$C(3)$ & $-0.160000$     & $1$ \\
\hline
\end{tabular}}
\caption{Parameters of the tight-binding model.}
\label{tab:TB}
\end{table}

To compute the induced magnetic field due to an external magnetic field, we calculate the energy difference between two perturbed systems: the molecule placed in a uniform magnetic field with and without a small test dipole, located in space where we intend to calculate the NICS. When applying a uniform magnetic field $\mathbf{H}$, the vector potential at a given point in space $\mathbf{r}$ is given by:
\begin{equation}
    \mathbf{A}_0(\mathbf{r}) = - \frac{1}{2}\mathbf{r}\times\mathbf{H}
\end{equation}
whereas in the case of a point magnetic dipole $\mathbf{m}$ located at $\mathbf{r}_m$ the vector potential can be written as:
\begin{equation}
    \mathbf{A}_m(\mathbf{r}) = \frac{\mu_o}{4\pi}\frac{\mathbf{m}\times(\mathbf{r}-\mathbf{r}_m)}{|\mathbf{r}-\mathbf{r}_m|^3}
    \textrm{.}
\end{equation}
Following the derivation of McWeeny~\citep{mcweeny1958}, the elements of the perturbed Hamiltonian are obtained as
\begin{equation}
    \mathcal{H}_{ij}^{'} =  \mathcal{H}_{ij} \exp{\left\{ \frac{\pi i e}{h c} [\mathbf{A}(\mathbf{r}_i)-\mathbf{A}(\mathbf{r}_j)]\cdot(\mathbf{r}_i+\mathbf{r}_j)\right \}}
    \mathrm{,}
\end{equation}
where $\mathbf{r}_i$ is the position of the atom $i$. The overlap matrix elements are modified similarly.

As an additional contribution to the TB model, we consider each carbon atom magnetically polarisable with a polarisability of $10.0$~\r{A}$^3$ for two carbon neighbours and $6.5$~\r{A}$^3$ for three carbon neighbours. We obtained these values from fitting the NICS$_{\rm DFT}$ values of benzene and naphthalene using the combination of dipole and perturbative TB model. The values of NICS calculated through the TB approach are designated as NICS$_{\rm TB}$ in the remainder of the article.

\section{Molecular size effects for $^{13}$C NMR shifts and NICS}

We start our discussion by examining molecular size effects on the $^{13}$C NMR chemical shifts using the results for the coronoids which constitute a consistent series of molecules. While size effects have already been studied in such compounds~\citep{Moran03,Thonhauser09,forse2014,Merlet15}, results have not been reported on such a large range of molecules. Comparison with experimental results is also limited to small coronoids. For coronene, our calculations yield chemical shift values of 124.2~ppm for the inner carbon atom, 129.8~ppm for the outer bridging, and 128.2~ppm for the protonated one. These values are found to be in good agreement with previous $^{13}$C NMR experimental studies~\citep{orendt2000,hughes1993,resing1987} which makes us confident that the chemical shifts calculated are reliable.

\iffalse
Relevant references:
Orendt2000: coronene exp: d$_{\rm protonated}$ = 123.5ppm d$_{\rm inner}$ = 121ppm, d$_{\rm bridging}$ = 124ppm (123, 120 and 123 for T=100K)
Hugues1993: coronene exp: d$_{protonated}$ = 124ppm, d$_{\rm bridging}$ = 126ppm, d$_{inner}$ = 120ppm
Resing1987: coronene exp: d$_{\rm protonated}$ = 122ppm, d$_{\rm bridging}$ = 125ppm, d$_{\rm inner}$ = 119ppm
\fi

\begin{figure}[ht!]
\begin{center}
\includegraphics[scale=0.8]{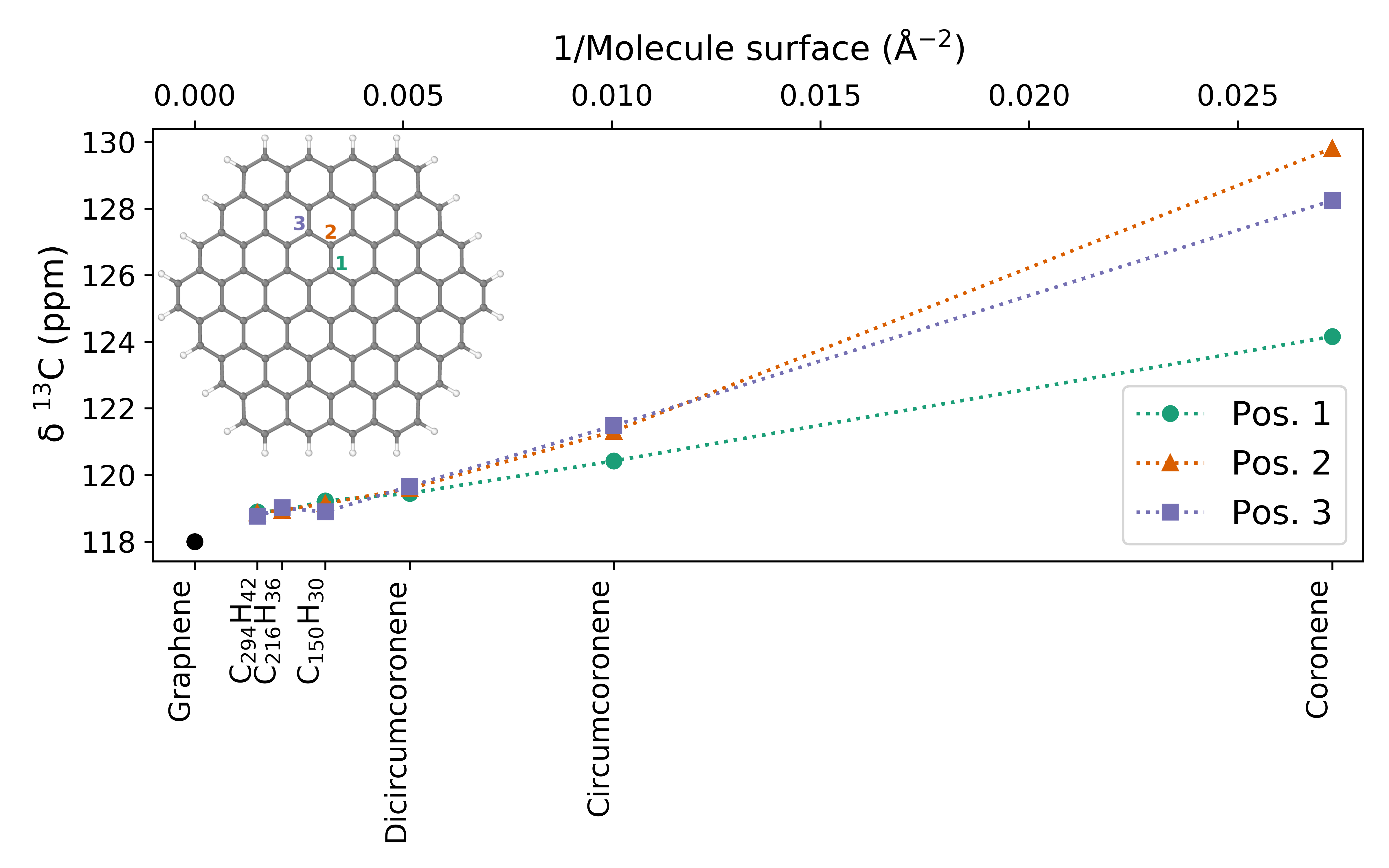}
\end{center}
\caption{Evolution of the $\mathrm{^{13}C}$ NMR shift of three equivalent positions with respect to the inverse molecule size. The value for graphene is taken from Ref.~\citep{Thonhauser09}.}
\label{fig:size-convergence}
\end{figure}

In Fig.~\ref{fig:size-convergence} we present the evolution of the $\mathrm{^{13}C}$ chemical shifts for three equivalent positions in the series of coronenes. We observe that, as the molecule size increases, the chemical shifts for all equivalent positions decrease and eventually converge towards a single value which is found to be close to the 118.0~ppm value reported by Thonhauser~et~al. for bulk graphene~\citep{Thonhauser09}. The converged values for molecules larger than dicircumcoronene indicate that, as far as chemical shifts are concerned, an atom can be considered to be in the bulk if it is found at a distance of at least three hexagonal rings from the molecule edge.

\begin{figure}[ht!]
\begin{center}
\includegraphics[scale=0.6]{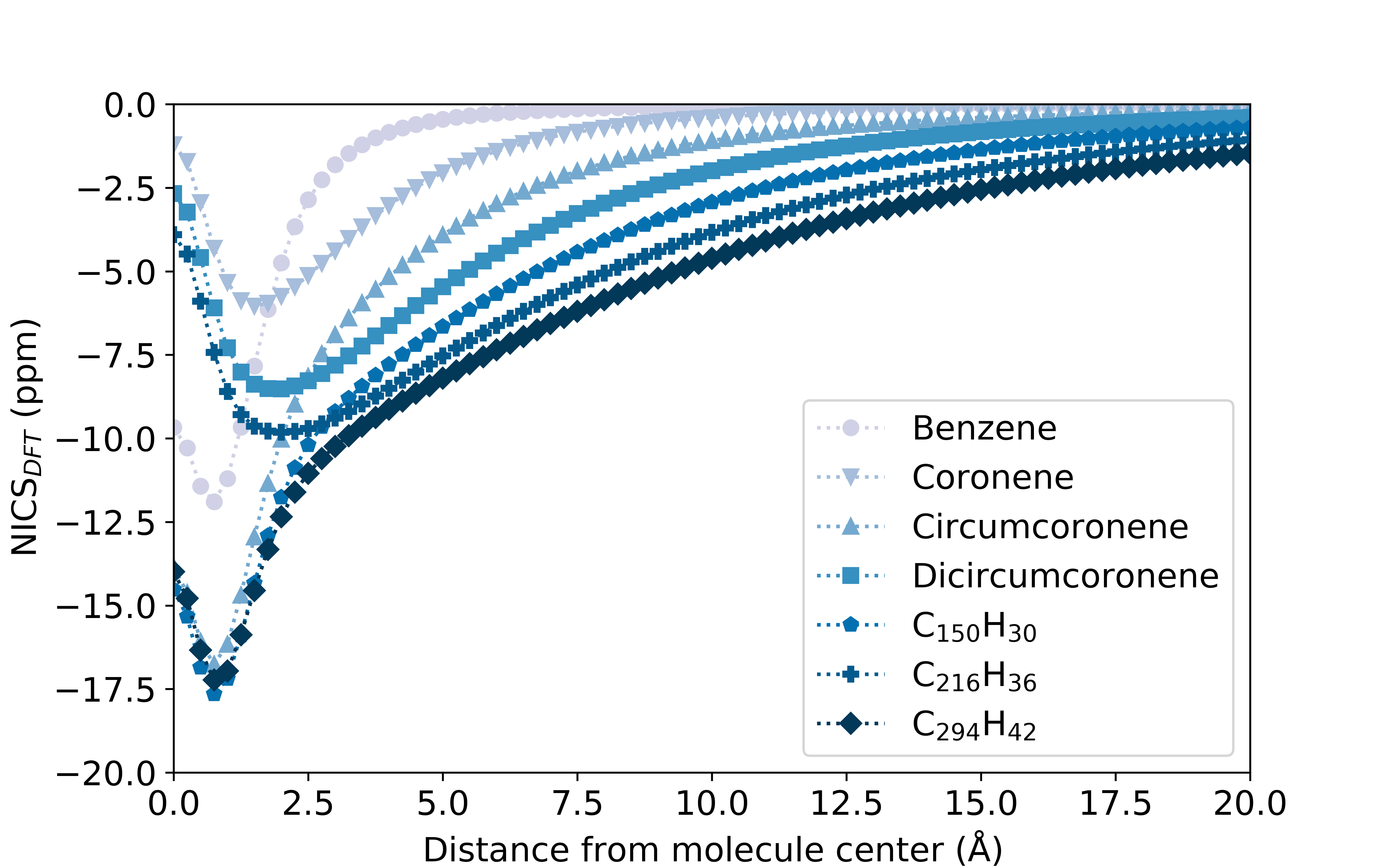}
\end{center}
\caption{DFT Isotropic NICS for the coronoid series, calculated on a line perpendicular to the molecule and going through the molecule center.}
\label{fig:NICS-line}
\end{figure}

We now turn to the discussion of the NICS, which are the chemical shift changes anticipated for a probe species due to ring current effects. We again focus on molecules in the coronoid series which have often been used as simple models to describe the chemical shifts observed for species adsorbed in disordered porous carbons~\citep{forse2014,Forse15b,Xing14,Merlet15}. In Fig.~\ref{fig:NICS-line} we present the calculated isotropic NICS values on a line that is perpendicular to the plane of the molecule and goes through its center. The evolution of the isotropic NICS with respect to the distance from the molecules is found to be in good agreement with previous theoretical works on the same systems~\citep{stanger2006,ciesielski2009,forse2014,charistos2019}. 

Close to the molecule, the NICS exhibit two distinct behaviours alternating between molecules showing a minimum isotropic NICS around 0.75~\r{A} ("benzene-like") and the ones that have a minimum between 1.5-2.2~\r{A} ("coronene-like"). This even-odd behavior has been previously reported by Hajgat\'o~et~al. who used energetic considerations to explain the presence of a strong aromatic central hole at the center of coronenes with an even number of ring layers and a weaker benzenoid central hole for the ones with an odd number of layers~\citep{hajgato2006}. Similarly, Sakamoto~et~al. discuss this effect as the result of delocalized aromaticity due to the presence of multiple Clar formulas for odd-numbered molecules~\citep{sakamoto2014}.

For distances larger than around 3.0~\r{A}, the isotropic NICS exhibit a clear trend, whereby the magnitude of the NICS increases with the molecule size. Similarly to the $^{13}$C chemical shifts, at these distances we observe a convergence of the values for the larger molecules, i.e. C$_{294}$H$_{42}$ and C$_{216}$H$_{36}$ show closer NICS than benzene and coronene (see Figure~S3). It is worth noting that the molecule size also affects the distance where the NICS values converge to zero, with this distance at approximately 7~\r{A} for benzene and above 50~\r{A} for C$_{294}$H$_{42}$ (see Fig.~S3 in Supplementary Material).

\section{Calculating the NICS efficiently}

\subsection{Using a classical dipolar model}

As mentioned in the introduction, NICS are often calculated to interpret the NMR spectra of ions or molecules adsorbed in porous carbons or in proximity of PAHs. As such, it would be very useful to be able to predict such chemical shifts at a low computational cost for a wide range of molecules. To this aim, we first test a classical dipolar model which has been proposed as an alternative for calculating ring currents~\citep{carrington1967}. This very crude model has the advantage of being very fast as one just needs to identify the centre of mass of the rings and calculate the distances between these centres of masses and the positions at which we evaluate the NICS. 

\begin{figure}[ht!]
\begin{center}
\includegraphics[scale=0.5]{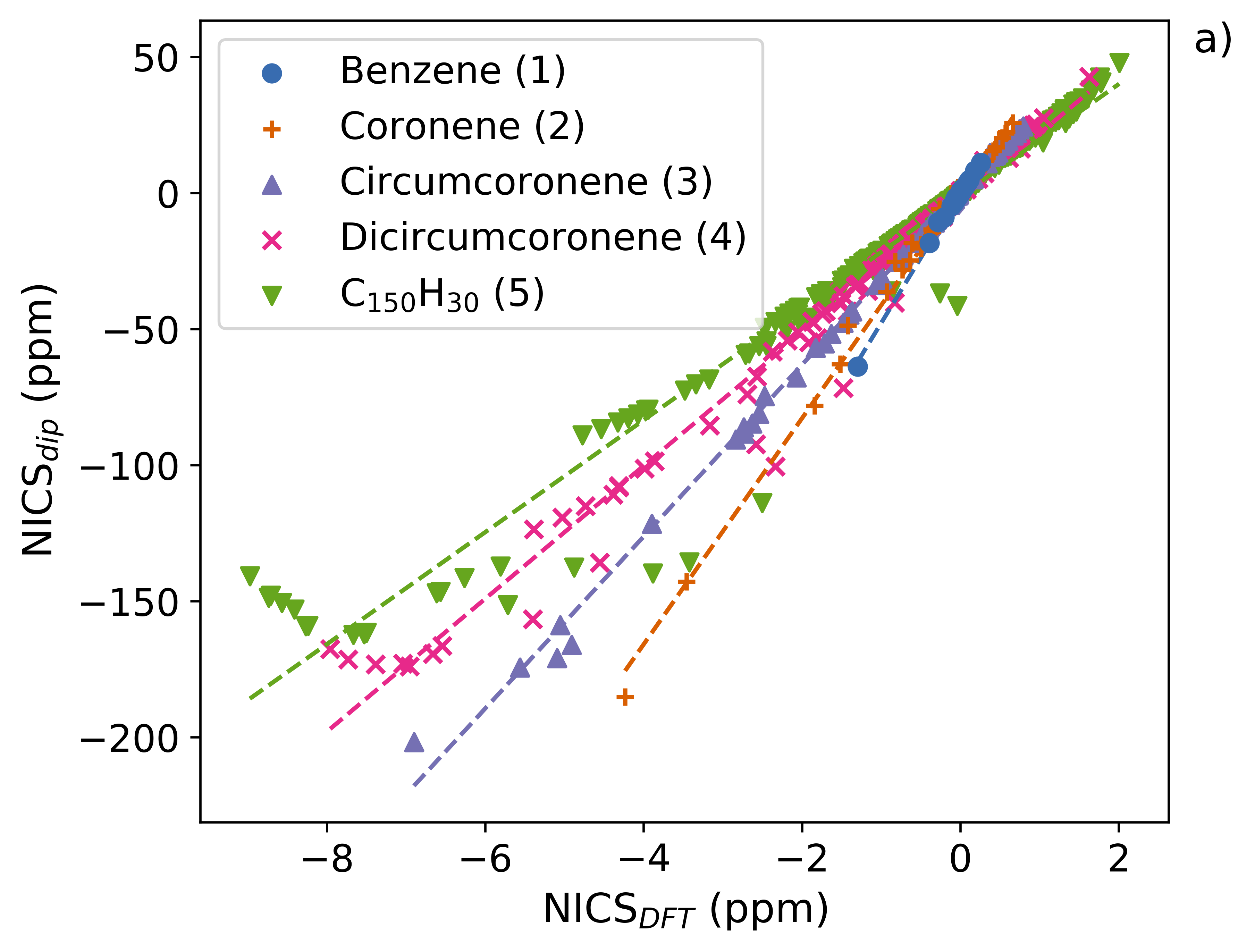}
\hskip 15pt
\includegraphics[scale=0.5]{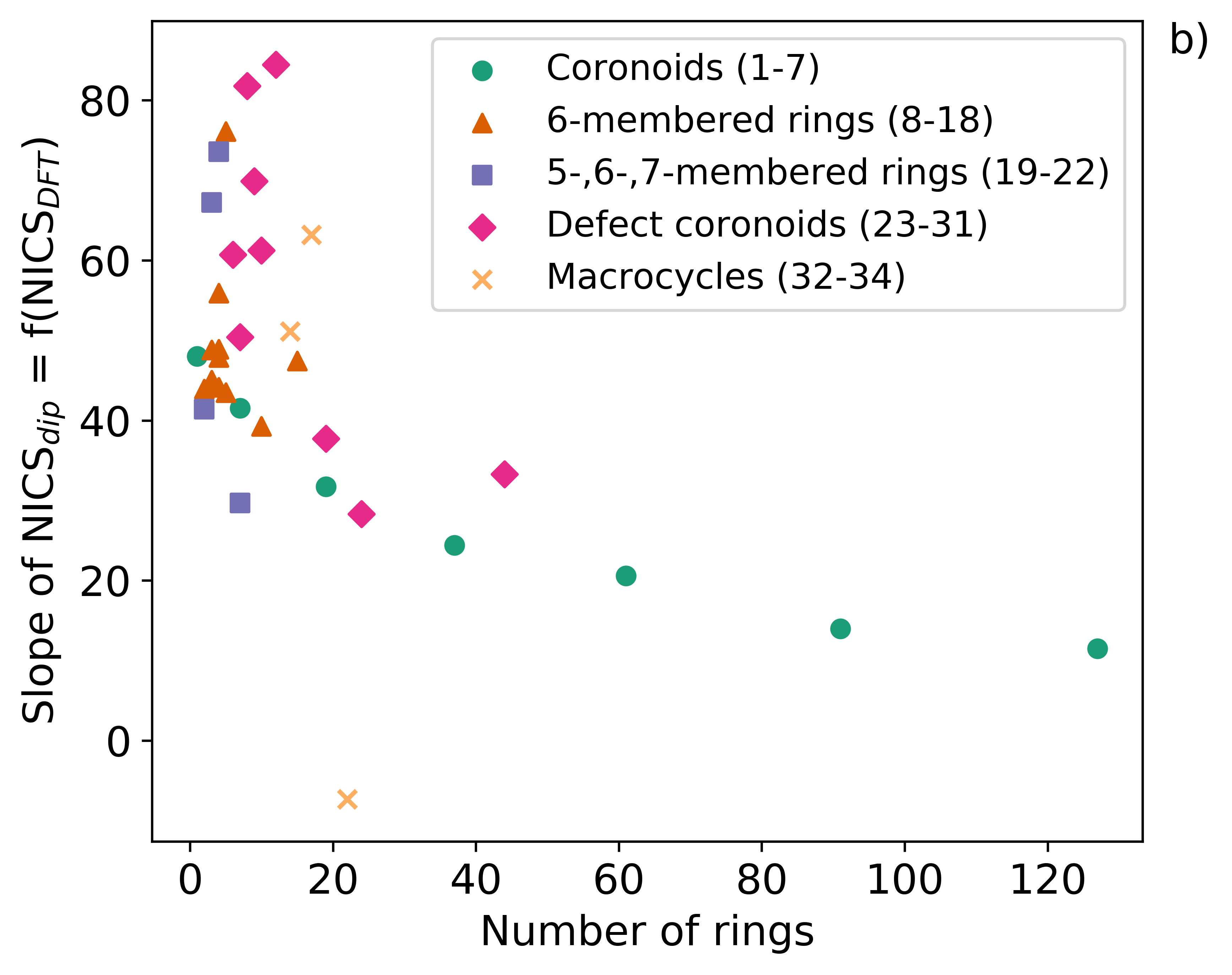}
\end{center}
\caption{a) Comparison of the NICS values calculated from a classical dipolar model with the values obtained by DFT for the coronene series and a distance cutoff of 3~\r{A}. Dashed lines indicate the corresponding linear fits. b) Slopes of the linear fits with respect to the number of rings for all molecules in the study. The molecule number according to Fig.~\ref{fig:coronenes} is given in parentheses. The results for a 5~\r{A} cutoff are given in Supplementary Material.}
\label{fig:Comp-dip-DFT}
\end{figure}

Fig.~\ref{fig:Comp-dip-DFT}a) shows the comparison between the NICS predicted via the dipolar model, NICS$_{\rm dip}$, and the values calculated using DFT, NICS$_{\rm DFT}$, for benzene and the coronene series. The NICS are calculated for the same grid points in both methods and a cutoff is applied: only points more than 3~\r{A} away from all carbon atoms are considered. Results obtained for a larger cutoff of 5~\r{A} are shown in Fig.~S5. Our choice for these cutoff values of 3~\r{A}, respectively 5~\r{A}, follows the fact that alkali ions (Li$^+$, Na$^+$, etc...), respectively molecular ions (PF$_6^-$, BF$_4^-$, etc...) will in most cases show free energy minima for adsorption at distances slightly larger than these values~\citep{Ivanistsev16,Gomez-Gonzalez18b,Merlet13c,Fedorov14} so that shorter distances are usually irrelevant for the description of NMR parameters of adsorbed species while it is crucial that the model is precise for larger distances. 

For the five molecules shown here, there is a clear correlation between the NICS calculated from the two methods but there are also many outliers. More precisely, the accuracy of the prediction decreases with the increase of the magnitude of the chemical shift. This is not surprising as larger chemical shifts correspond to smaller probe distances (see Fig.~\ref{fig:NICS-line}) and the dipolar model is only valid for relatively large distances where an effective induced magnetic field is a more valid approximation. Indeed, the results for the 5~\r{A} cutoff show much less outliers (see Fig.~S5). It is worth noting that the NICS$_{\rm dip}$ values are much larger than the NICS$_{\rm DFT}$ values. This is a result of the fact that we used the magnetic susceptibilities of graphite. The magnetic susceptibilities calculated by Schulman and Dish show that $\chi_\parallel$ is two orders of magnitude (respectively one order of magnitude) smaller for benzene (respectively coronene) compared to graphite~\citep{Schulman97}. 

Since the goal of this work is to find an efficient way to calculate the NICS without manual adjustment of the parameters, it is not desirable to include experimental or DFT based estimations of the magnetic susceptibility in the dipolar model approach. Moreover, in this model, a perfect additivity of the ring contributions to the total NICS is assumed, while Facelli has shown that this is not valid~\citep{Facelli06}. To explore the possibility of including these effects in a semi-empirical way, we check if the variation of the slope of the correlation plot shows a predictable evolution with the number of rings. A plot showing the slopes calculated for all PAHs studied here is shown in Fig.~\ref{fig:Comp-dip-DFT}b). As expected from the molecular size effects discussed in the previous section there is a clear trend for benzene and the coronene series. The other small molecules show large, nonintuitive, variations which prevent using the dipolar model without further, probably complex, refinement.    

Overall, due to its inability to predict chemical shifts for distances close to the PAHs and the lack of a simple way to include the variation of the magnetic susceptibility, the dipolar model seems insufficient to predict accurately the NICS for a wide range of molecules.

\subsection{Using a perturbative tight-binding model}

\begin{figure}[ht!]
\begin{center}
\includegraphics[scale=0.8]{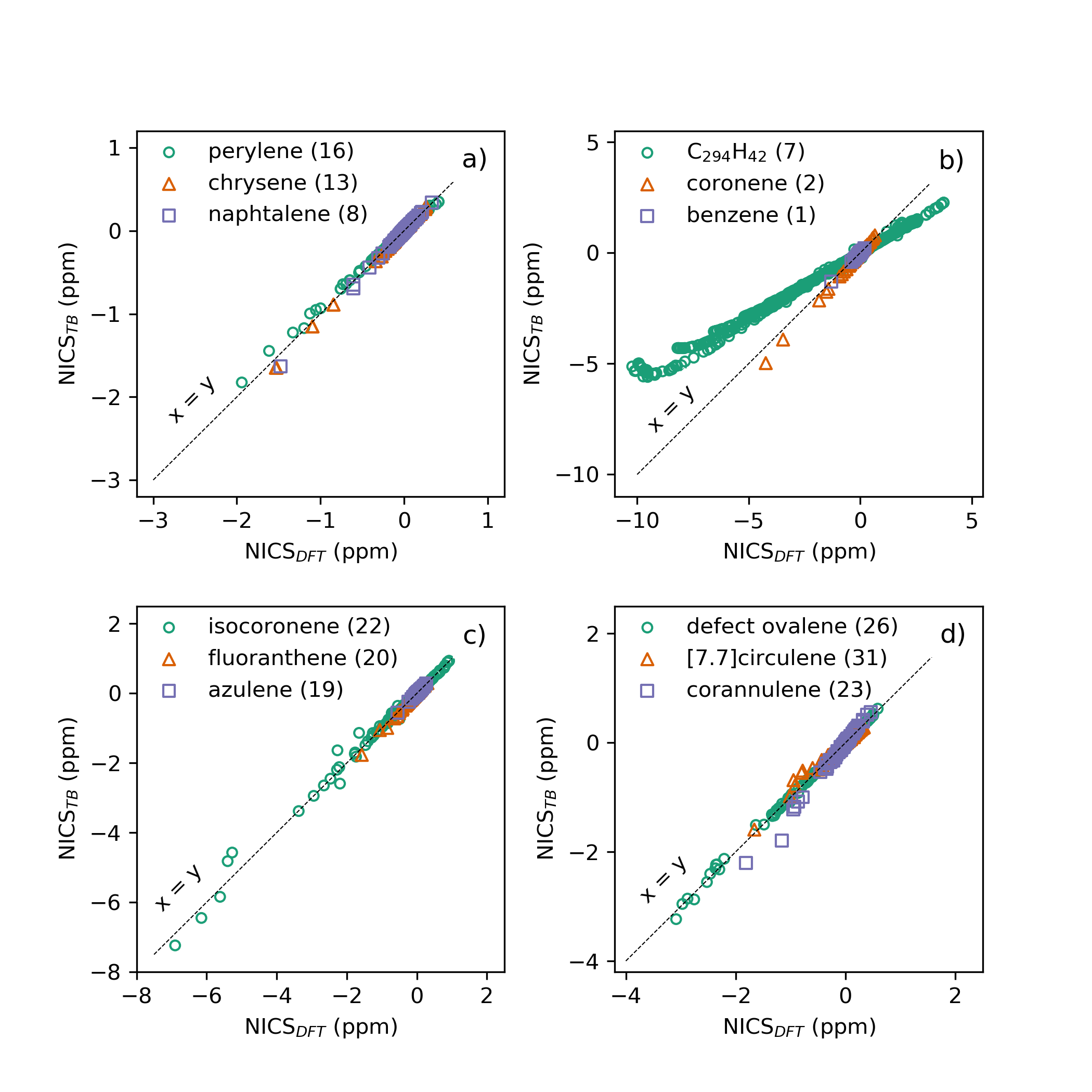}
\end{center}
\caption{Comparison of isotropic NICS calculated through tight-binding and DFT methods on a grid around a series of aromatic hydrocarbons with a cutoff of 3~\r{A}. The molecule number according to Fig.~\ref{fig:coronenes} is given in parentheses. The correlations for all studied molecules are given as Supplementary Material.}
\label{fig:correlations}
\end{figure}

Following the observation that a simple dipolar model is insufficient to accurately predict the NICS, we turn to a more original perturbative tight-binding model approach, which excitingly provides much better agreement with DFT values. As in the case of the dipolar model, we test this approach using a minimum distance cutoff of 3~\r{A} and a larger cutoff set to 5~\r{A}. Fig.~\ref{fig:correlations} shows the comparison between the tight-binding model results and the values calculated using DFT for a subset of molecules and a cutoff of 3~\r{A}. Except for C$_{294}$H$_{42}$, a large coronoid, the results are extremely good with a slope close to unity. The results for the complete set of molecules are given in Supplementary Material. NICS$_{\rm TB}$ and NICS$_{\rm DFT}$ values are very well correlated in all cases, i.e. even for some non planar molecules, which is surprising considering the minimal basis set adopted here consisting exclusively of $p$ orbitals perpendicular to the mirror plane of the molecule.    

To fully characterize the performance of the tight-binding model, we calculate average slopes, correlation coefficients and average maximal errors for all the molecules and the two cutoffs. The results averaged for groups of molecules are summarized in Table~\ref{tab:avg_correl} while results for each individual molecule are available in Supplementary Material. The results for the 3~\r{A} cutoff already show a very good agreement between the tight-binding model and the DFT calculations, with the correlation slopes being very close to unity for the majority of the molecules under study.  For the longer cutoff of 5~\r{A}, the results are even more accurate, showing the model's improved performance at intermediate to long distances. Overall, the average maximum errors for the isotropic NICS for all molecules in the set were found to be respectively 0.9~ppm and 0.5~ppm for the 3~\r{A} and 5~\r{A} cutoffs, which is close to the experimental uncertainty.   

\begin{table} [ht!]
{\setlength{\tabcolsep}{0.21cm}
\begin{tabular}{l c c c c c c c c c}
\hline
& \multicolumn{4}{c}{r$_{\rm cut}$ = 3~\r{A}} & & \multicolumn{4}{c}{r$_{\rm cut}$ = 5~\r{A}} \\
 & slope & R$^2$ & Err$_{\rm max,iso}$ & Err$_{\rm max,ZZ}$ & & slope & R$^2$ & Err$_{\rm max,iso}$ & Err$_{\rm max,ZZ}$ \\
 & & & (ppm) & (ppm) & & & & (ppm) & (ppm) \\
\hline
Coronoids           & 0.98 & 0.9991 & 5.1 & 15.4 & & 0.99 & 0.9996 & 3.9 & 11.9 \\
6-membered rings      & 1.07 & 0.9993 & 1.0 & 2.6 & & 1.07 & 0.9995 & 0.4 & 1.0 \\
5,6,7-membered rings  & 1.05 & 0.9970 & 0.7 & 2.3 & & 1.04 & 0.9993 & 0.1 & 0.5 \\
Defect coronoids    & 1.07 & 0.9927 & 1.8 & 6.0 & & 1.06 & 0.9931 & 0.8 & 2.6 \\
Macrocycles         & 1.43 & 0.9908 & 10.8 & 32.7 & & 1.45 & 0.9971 & 6.4 & 19.3 \\
\hline
\end{tabular}}
\caption{Average slopes and correlation coefficients for the fits between the isotropic values of NICS$_{\rm TB}$ and NICS$_{\rm DFT}$ for different groups of PAHs, as well as the average maximal error encountered for the isotropic NICS and the ZZ component of the NICS tensor. The results are given for two different cutoff values, 3~\r{A} and 5~\r{A}.}
\label{tab:avg_correl}
\end{table}

The molecules for which the tight-binding model seems the least appropriate are C$_{294}$H$_{42}$ (\textbf{7}) and chrysaorole (\textbf{32}). For these molecules, the correlation slope deviates from unity and the tight-binding model does not predict accurately the isotropic NICS for some grid points and reaches a maximal error of 5.1~ppm for C$_{294}$H$_{42}$ and of 10.8~ppm for chrysaorole. It is worth noting that for coronoids larger than circumcoronene, the slope of the correlation decreases systematically. This could be due to the inability of our model to capture the change in magnetic susceptibility, $\mu_0$, with the molecular size. Indeed, in the current tight-binding model, the magnetic susceptibility is considered via a constant polarisability which depends only on the number of neighbouring carbons while the size effects described in previous sections suggest that $\mu_0$ should depend on the molecular size. This could be an avenue for further refinement of the model. 

\begin{figure}[ht!]
\begin{center}
\includegraphics[scale=0.8]{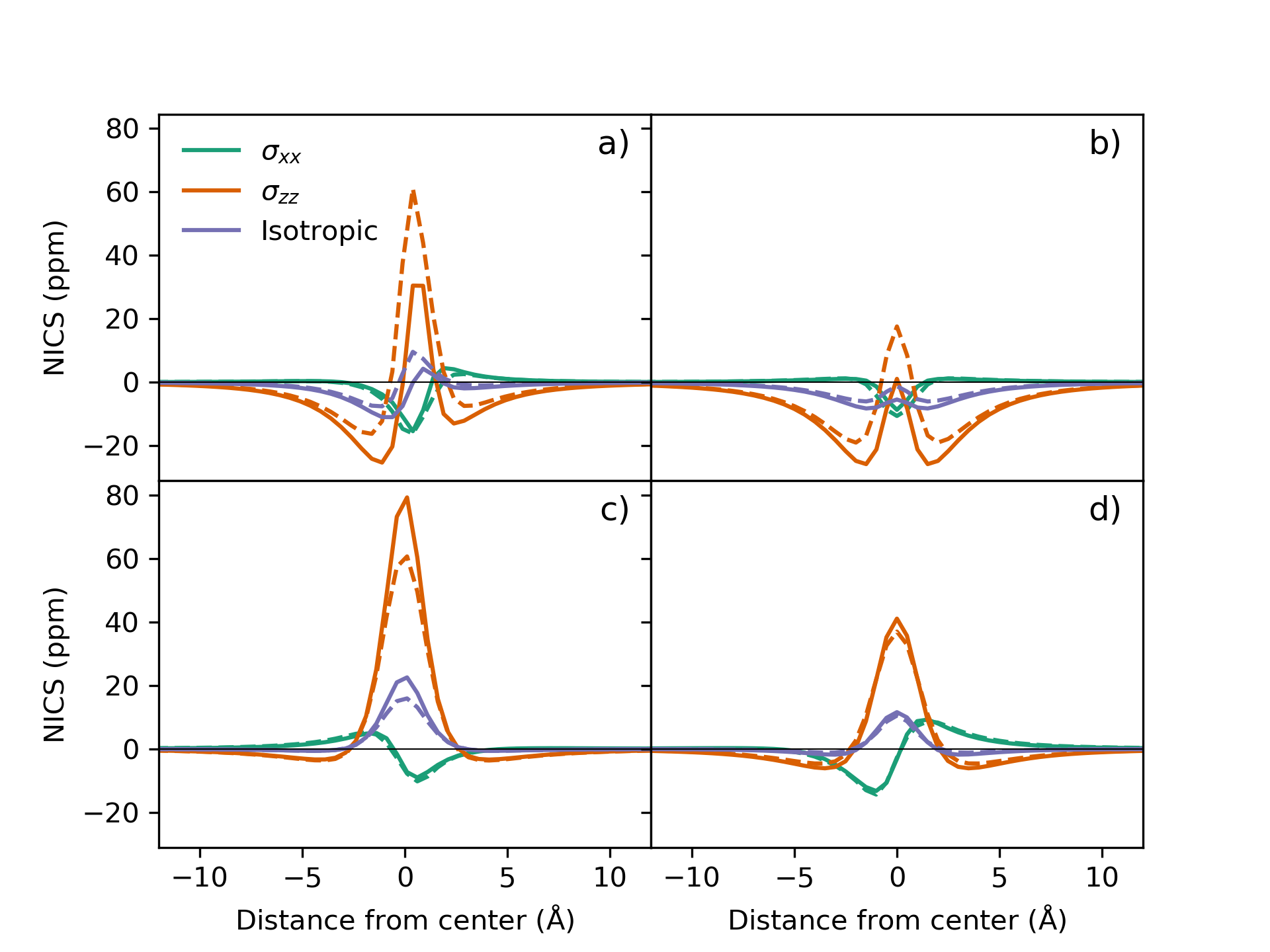}
\end{center}
\caption{$xx$, $zz$ and isotropic components of the NICS tensor calculated on a line perpendicular to the central ring for a) coronene (\textbf{2}), b) corannulene (\textbf{23}), c) [7]circulene (\textbf{24}) and d) [8]circulene (\textbf{25}). Solid lines are the values obtained by the TB model and dashed lines are DFT calculations.}
\label{fig:central-ring}
\end{figure}

Concerning the individual tensor elements, the largest errors are usually observed for the $zz$ component as can be seen in Tables~\ref{tab:avg_correl}, S1 and S2. This can be visualized more precisely in Fig.~\ref{fig:central-ring}, in which we compare the tight-binding results for the $\sigma_{xx}$, $\sigma_{zz}$, and isotropic NICS of a series of simple cyclic PAHs, having a different central ring size, to the values obtained using DFT. In this case, the NICS were calculated on a line going through the center of mass of the molecule. Concerning the DFT values for the isotropic NICS, we find a good agreement with the previous works by Forse~et~al.~\citep{forse2014} for both corannulene and coronene, as well as the ones by Ciesielski~et~al. for the latter~\citep{ciesielski2009}. However, we do observe a slight discrepancy when comparing our $zz$ component value close to the ring center with the one recently presented by Charistos~et~al.~\citep{charistos2019}. As far as the results from the tight-binding model are concerned, we observe a very good agreement with the DFT values in all cases, especially for distances greater than 3~\r{A}, whereas quantitative discrepancies mostly concern the $zz$ component close to the molecule centers. 

Overall, the results suggest that the simple tight-binding model presented herein can predict NICS values for a variety of aromatic hydrocarbons with a good accuracy and a much lower computational cost than DFT. Indeed, computational times shown in Table~\ref{tab:timings} demonstrate that the current tight-binding model, while not yet fully optimised, is one to two orders of magnitude faster than DFT. The additional cost of the tight-binding model compared to the dipolar model is totally justified by its superior accuracy. We have shown that the tight-binding model is not only restricted to molecules solely consisting of 6-membered rings, but can also be efficiently applied to systems with 5- and 7-membered ones. Moreover, although it was not initially conceived for such cases, we demonstrated that the tight-binding model can also accurately predict the NICS for some simple non-planar molecules. 

\begin{table} [ht!]
{\setlength{\tabcolsep}{0.25cm}
\begin{tabular}{l c c c }
\hline
  & t$_{DFT}$ (s) & t$_{TB}$ (s) & t$_{dip}$ (s) \\
\hline
Benzene (\textbf{1}) &  6.7$\times10^{-2}$ & 3.3$\times10^{-3}$ & $<10^{-5}$ \\
Pentacene (\textbf{11}) & 6.2$\times10^{-1}$ & 5.7$\times10^{-3}$ & $<10^{-5}$ \\
Coronene (\textbf{2}) & 8.6$\times10^{-1}$ & 7.7$\times10^{-3}$ & $<10^{-5}$ \\
Chrysaorole (\textbf{32}) & $3.9\times10^{-1}$ & 7.4$\times10^{-3}$ & $<10^{-5}$ \\
C$_{150}$H$_{30}$ (\textbf{5}) & 3.1 & 1.3$\times10^{-2}$ & $<10^{-5}$ \\
C$_{294}$H$_{42}$ (\textbf{7}) & 6.9 & 1.4$\times10^{-2}$ & $<10^{-5}$ \\
\hline
\end{tabular}}
\caption{Computational times (per NICS grid point and per carbon atom) for DFT, tight-binding model, and dipolar model calculations, for a representative subset of molecules. All results have been obtained on 2.3~GHz Intel\textsuperscript{\textregistered} Skylake 6140 processors (parallel processes for DFT).}
\label{tab:timings}
\end{table}

There are still cases where the simple tight-binding model deviates from DFT calculations, as presented above. Moreover, the use of just one $p$ orbital per carbon atom prohibits the model from being applied to more complex environments. Such cases could be molecules with an important curvature in their structure, the presence of sp$^3$ carbons, as well as charged species. The correct prediction of the NICS tensor in these cases would be an important step towards calculations for periodic structures of disordered carbons. These issues will be addressed in the future by refining the tight-binding model and employing an extended basis set for the calculations.

\section{Conclusion}

In this work, we report on $^{13}$C and nucleus independent chemical shifts for a range of PAHs. DFT calculations allowed us to investigate molecular size effects and have been used as a reference for the evaluation of two computationally cheap methods for the calculation of NICS. It was shown that, in the coronoid series, the $^{13}$C chemical shifts converge towards the graphene values, while the NICS show an odd-even behavior with the number of ring layers. The first model tested for the cheap computation of NICS was a classical dipolar model. While NICS values calculated with such a model show a clear correlation with DFT results at relatively large distances from the considered molecules, it is not valid at small distances and it does not allow for a straightforward inclusion of the magnetic susceptibility. We then proposed a simple tight-binding model with a basis set consisting of a single $p$ orbital per cabon atom. This model gives strikingly good results for the set of 34 molecules studied here, including some non planar cases. This is very promising and demonstrates that such an approach could be suitable for the description of more complex carbon structures.

\section*{Supplementary Material}

See supplementary material for a comparison of chemical shifts calculated using DFT with different basis sets, a visualization of the grid points for which NICS are determined, the values of isotropic NICS for coronoids at a distance of 3~\r{A} from the molecule center, a visualization of the dipolar model geometrical parameters, and detailed results for the dipolar and tight-binding models for the two cutoffs selected.

\section*{Acknowledgements}

This project has received funding from the European Research Council (ERC) under the European Union’s Horizon 2020 research and innovation program (grant agreement no. 714581). This work was granted access to the HPC resources of CALMIP supercomputing center under the allocation P19003 and of TGCC under the allocation 2019-A0070911061 made by GENCI. A.B.P. acknowledges support from the Collaborative Computational Project for NMR Crystallography (CCP-NC) and UKCP Consortium, both funded by the Engineering and Physical Sciences Research Council (EPSRC) under Grant Nos. EP/M022501/1 and EP/P022561/1, respectively. Some of the calculations were run using the STFC Scientific Computing Department’s SCARF cluster. C.J.P. is supported by the Royal Society through a Royal Society Wolfson Research Merit Award and the EPSRC through Grant No. EP/P022596/1.

\newpage

%\bibliographystyle{unsrt}
%\bibliography{references}

\begin{thebibliography}{10}

\bibitem{Blanc13}
F.~Blanc, M.~Leskes, and C.~P. Grey.
\newblock \emph{In situ} solid-state {NMR} spectroscopy of electrochemical
  cells: {B}atteries, supercapacitors, and fuel cells.
\newblock {\em Acc. Chem. Res.}, 46:1952--1963, 2013.

\bibitem{Griffin16}
J.~M. Griffin, A.~C. Forse, and C.~P. Grey.
\newblock Solid-state {NMR} studies of supercapacitors.
\newblock {\em Solid State Nucl. Mag.}, 74-75:16 -- 35, 2016.

\bibitem{wang2011}
H.~Wang, T.~K.-J. K\"{o}ster, N.~M. Trease, J.~S{\'e}galini, P.-L. Taberna,
  P.~Simon, Y.~Gogotsi, and C.~P. Grey.
\newblock Real-time {NMR} studies of electrochemical double-layer capacitors.
\newblock {\em J. Am. Chem. Soc.}, 133:19270--19273, 2011.

\bibitem{shen2008}
Y.~Shen, O.~Lange, F.~Delaglio, P.~Rossi, J.~M. Aramini, G.~Liu, A.~Eletsky,
  Y.~Wu, K.~K. Singarapu, A.~Lemak, et~al.
\newblock Consistent blind protein structure generation from {NMR} chemical
  shift data.
\newblock {\em P. Natl. Acad. Sci.}, 105:4685--4690, 2008.

\bibitem{Cavanagh_book}
J.~Cavanagh, W.~J. Fairbrother, A.~G. Palmer~III, and N.~J. Skelton.
\newblock {\em Protein {NMR} spectroscopy, principles and practice}.
\newblock Academic Press, Inc., 1996.

\bibitem{Marion13}
D.~Marion.
\newblock An introduction to biological {NMR} spectroscopy.
\newblock {\em Mol. Cell. Proteomics}, 12:3006--3025, 2013.

\bibitem{Youngman18}
R.~Youngman.
\newblock {NMR} spectroscopy in glass science: {A} review of the elements.
\newblock {\em Materials}, 11:476, 2018.

\bibitem{stebbins1997}
J.~F. Stebbins and Z.~Xu.
\newblock {NMR} evidence for excess non-bridging oxygen in an aluminosilicate
  glass.
\newblock {\em Nature}, 390:60, 1997.

\bibitem{schleder2019}
G.~R Schleder, A.~C.~M. Padilha, C.~M. Acosta, M.~Costa, and A.~Fazzio.
\newblock From {DFT} to machine learning: {R}ecent approaches to materials
  science--a review.
\newblock {\em J. Phys.: Mater.}, 2:032001, 2019.

\bibitem{haigh1979}
C.~W. Haigh and R.~B. Mallion.
\newblock Ring current theories in nuclear magnetic resonance.
\newblock {\em Prog. Nucl. Mag. Res. Sp.}, 13:303--344, 1979.

\bibitem{kvasnicka1991}
V.~Kvasni{\^c}ka.
\newblock An application of neural networks in chemistry. {P}rediction of
  $^{13}${C} {NMR} chemical shifts.
\newblock {\em J. Math. Chem.}, 6:63--76, 1991.

\bibitem{meiler2000}
J.~Meiler, R.~Meusinger, and M.~Will.
\newblock Fast determination of {$^{13}$C NMR} chemical shifts using artificial
  neural networks.
\newblock {\em Journal Chem. Inf. Comp. Sci.}, 40:1169--1176, 2000.

\bibitem{paruzzo2018}
F.~M. Paruzzo, A.~Hofstetter, F.~Musil, S.~De, M.~Ceriotti, and L.~Emsley.
\newblock Chemical shifts in molecular solids by machine learning.
\newblock {\em Nature commun.}, 9:4501, 2018.

\bibitem{gerrard2019}
W.~Gerrard, L.~A. Bratholm, M.~Packer, A.~J. Mulholland, D.~R. Glowacki, and
  C.~P. Butts.
\newblock {IMPRESSION}-prediction of {NMR} parameters for 3-dimensional
  chemical structures using machine learning with near quantum chemical
  accuracy.
\newblock {\em Chem. Sci.}, 2019.

\bibitem{jonas2019}
E.~Jonas and S.~Kuhn.
\newblock Rapid prediction of {NMR} spectral properties with quantified
  uncertainty.
\newblock {\em J. Cheminformatics}, 11:1--7, 2019.

\bibitem{cuny2016}
J.~Cuny, Y.~Xie, C.~J. Pickard, and A.~A. Hassanali.
\newblock \emph{Ab initio} quality {NMR} parameters in solid-state materials
  using a high-dimensional neural-network representation.
\newblock {\em J. Chem. Theory Comput.}, 12:765--773, 2016.

\bibitem{chaker2019}
Z.~Chaker, M.~Salanne, J.-M. Delaye, and T.~Charpentier.
\newblock {NMR} shifts in aluminosilicate glasses via machine learning.
\newblock {\em Phys. Chem. Chem. Phys.}, 21:21709--21725, 2019.

\bibitem{Zhao06}
X.~S Zhao.
\newblock Novel porous materials for emerging applications.
\newblock {\em J. Mater. Chem.}, 16:623--625, 2006.

\bibitem{Xing14}
Y.-Z. Xing, Z.-X. Luo, A.~Kleinhammes, and Y.~Wu.
\newblock Probing carbon micropore size distribution by nucleus independent
  chemical shift.
\newblock {\em Carbon}, 77:1132--1139, 2014.

\bibitem{Forse15b}
A.~C. Forse, C.~Merlet, P.~K. Allan, E.~K. Humphreys, J.~M. Griffin, M.~Aslan,
  M.~Zeiger, V.~Presser, Y.~Gogotsi, and C.~P. Grey.
\newblock New insights into the structure of nanoporous carbons from {NMR},
  {R}aman, and pair distribution function analysis.
\newblock {\em Chem. Mater.}, 27:6848--6857, 2015.

\bibitem{Griffin14}
J.~M. Griffin, A.~C. Forse, H.~Wang, Nicole~M. Trease, P.-L. Taberna, P.~Simon,
  and C.~P. Grey.
\newblock Ion counting in supercapacitor electrodes using {NMR} spectroscopy.
\newblock {\em Faraday Discuss.}, 176:49, 2014.

\bibitem{Forse17}
A.~C. Forse, J.~M. Griffin, C.~Merlet, J.~Carreteo-Gonzalez, A.-R.~O. Raji,
  N.~M. Trease, and C.~P. Grey.
\newblock Direct observation of ion dynamics in supercapacitor electrodes using
  \emph{in situ} diffusion {NMR} spectroscopy.
\newblock {\em Nature Ener.}, 2:16216, 2017.

\bibitem{Borchardt18}
L.~Borchardt, D.~Leistenschneider, J.~Haase, and M.~Dvoyashkin.
\newblock Revising the concept of pore hierarchy for ionic transport in carbon
  materials for supercapacitors.
\newblock {\em Adv. Ener. Mater.}, 8:1800892, 2018.

\bibitem{nandy2018}
A.~Nandy, A.~C. Forse, V.~J. Witherspoon, and J.~A. Reimer.
\newblock {NMR} spectroscopy reveals adsorbate binding sites in the
  metal--organic framework {UiO-66 (Zr)}.
\newblock {\em J. Phys. Chem. C}, 122:8295--8305, 2018.

\bibitem{zhang2018}
Y.~Zhang, B.~E.~G. Lucier, Michael Fischer, Zhehong Gan, Paul~D Boyle, Bligh
  Desveaux, and Yining Huang.
\newblock A multifaceted study of methane adsorption in metal--organic
  frameworks by using three complementary techniques.
\newblock {\em Chem: Eur. J.}, 24:7866--7881, 2018.

\bibitem{white1992}
J.~L. White, L.~W. Beck, and J.~F. Haw.
\newblock Characterization of hydrogen bonding in zeolites by proton
  solid-state {NMR} spectroscopy.
\newblock {\em J. Am. Chem. Soc.}, 114:6182--6189, 1992.

\bibitem{forse2014}
A.~C. Forse, J.~M. Griffin, V.~Presser, Y.~Gogotsi, and C.~P. Grey.
\newblock Ring current effects: {F}actors affecting the {NMR} chemical shift of
  molecules adsorbed on porous carbons.
\newblock {\em J. Phys. Chem. C}, 118:7508--7514, 2014.

\bibitem{resing1987}
H.~A. Resing and D.~L. VanderHart.
\newblock Coronene as a model of charcoal: {C}alibration of the carbon-13 {NMR}
  shift tensor to count carbon atoms at the plane edge.
\newblock {\em Z. Phys. Chem.}, 151:137--155, 1987.

\bibitem{Weisshoff02}
H.~Wei{\ss}hoff, A.~Prei{\ss}, I.~Nehls, T.~Win, and C.~M{\"u}gge.
\newblock Development of an {HPLC-NMR} method for the determination of {PAH}s
  in soil samples -- a comparison with conventional methods.
\newblock {\em Anal. Bioanal. Chem.}, 373:810--819, 2002.

\bibitem{harvey1988}
R.~G. Harvey and N.~E. Geacintov.
\newblock Intercalation and binding of carcinogenic hydrocarbon metabolites to
  nucleic acids.
\newblock {\em Acc. Chem. Res.}, 21:66--73, 1988.

\bibitem{umbuzeiro2008}
G.~A. Umbuzeiro, A.~Franco, M.~H. Martins, F.~Kummrow, L.~Carvalho, H.~H.
  Schmeiser, J.~Leykauf, M.~Stiborova, and L.~D. Claxton.
\newblock Mutagenicity and {DNA} adduct formation of {PAH}, nitro-{PAH}, and
  oxy-{PAH} fractions of atmospheric particulate matter from {S}ao {P}aulo,
  {B}razil.
\newblock {\em Mutat. Res. -- Gen. Tox. En.}, 652:72--80, 2008.

\bibitem{carrington1967}
A.~Carrington and A.~D. McLachlan.
\newblock {\em Introduction to magnetic resonance: {W}ith applications to
  chemistry and chemical physics}.
\newblock Harper and Row, 1967.

\bibitem{frisch2013}
M.~J. Frisch, G.~W. Trucks, H.~B. Schlegel, G.~E. Scuseria, M.~A. Robb, J.~R.
  Cheeseman, G.~Scalmani, V.~Barone, B.~Mennucci, G.~A. Petersson, et~al.
\newblock {Gaussian 09, Revision D. 01, 2013, Gaussian}.
\newblock {\em Inc., Wallingford CT}, 2013.

\bibitem{Moran03}
D.~Moran, F.~Stahl, H.~F. Bettinger, H.~F. Schaefer, and P.~v.~R. Schleyer.
\newblock Towards graphite: {M}agnetic properties of large polybenzenoid
  hydrocarbons.
\newblock {\em J. Am. Chem. Soc.}, 125:6746--6752, 2003.

\bibitem{becke1993}
A.~D. Becke.
\newblock Density-functional thermochemistry. {III}. {T}he role of exact
  exchange.
\newblock {\em J. Chem. Phys.}, 98:5648--5652, 1993.

\bibitem{Facelli06}
J.~C. Facelli.
\newblock Intermolecular shielding from molecular magnetic susceptibility. {A}
  new view of intermolecular ring current effects.
\newblock {\em Magn. Reson. Chem.}, 44:401--408, 2006.

\bibitem{gupta2005}
R.~R. Gupta, M.~D. Lechner, and B.~Mikhova.
\newblock {\em {NMR} Data for Carbon-13. Aromatic Compounds}.
\newblock Springer, 2005.

\bibitem{ganguli1941}
N.~Ganguli and K.~S. Krishnan.
\newblock Magnetic and other properties of the free electrons in graphite.
\newblock {\em P. Roy. Soc. Lond. A Mat.}, 177:168--182, 1941.

\bibitem{mcweeny1958}
R~McWeeny.
\newblock Ring currents and proton magnetic resonance in aromatic molecules.
\newblock {\em Mol. Phys.}, 1:311--321, 1958.

\bibitem{london1937}
F.~London.
\newblock Th{\'e}orie quantique des courants interatomiques dans les
  combinaisons aromatiques.
\newblock {\em J. Phys. Radium}, 8:397--409, 1937.

\bibitem{lazzeretti2000}
P.~Lazzeretti.
\newblock Ring currents.
\newblock {\em Prog. Nucl. Mag. Res. Sp.}, 36:1--88, 2000.

\bibitem{Thonhauser09}
T.~Thonhauser, D.~Ceresoli, and N.~Marzari.
\newblock {NMR} shifts for polycyclic aromatic hydrocarbons from
  first-principles.
\newblock {\em Int. J. Quantum Chem.}, 109:3336--3342, 2009.

\bibitem{Merlet15}
C.~Merlet, A.~C. Forse, J.~M. Griffin, D.~Frenkel, and C.~P. Grey.
\newblock Lattice simulation method to model diffusion and {NMR} spectra in
  porous materials.
\newblock {\em J. Chem. Phys.}, 142:094701, 2015.

\bibitem{orendt2000}
A.~M. Orendt, J.~C. Facelli, S.~Bai, A.~Rai, M.~Gossett, L.~T. Scott,
  J.~Boerio-Goates, R.~J. Pugmire, and D.~M. Grant.
\newblock Carbon-13 shift tensors in polycyclic aromatic compounds. 8. {A}
  low-temperature {NMR} study of coronene and corannulene.
\newblock {\em J. Phys. Chem. A}, 104:149--155, 2000.

\bibitem{hughes1993}
C.~D. Hughes, M.~H. Sherwood, D.~W. Alderman, and D.~M. Grant.
\newblock Chemical-shift-chemical-shift correlation spectroscopy in powdered
  solids.
\newblock {\em J. Magn. Reson., Ser. A}, 102:58--72, 1993.

\bibitem{stanger2006}
A.~Stanger.
\newblock Nucleus-independent chemical shifts ({NICS}): {D}istance dependence
  and revised criteria for aromaticity and antiaromaticity.
\newblock {\em J. Org. Chem.}, 71:883--893, 2006.

\bibitem{ciesielski2009}
A.~Ciesielski, T.~M. Krygowski, M.~K. Cyra{\'n}ski, M.~A. Dobrowolski, and
  J.-I. Aihara.
\newblock Graph-topological approach to magnetic properties of benzenoid
  hydrocarbons.
\newblock {\em Phys. Chem. Chem. Phys.}, 11:11447--11455, 2009.

\bibitem{charistos2019}
N.~D. Charistos, A.~Muñoz-Castro, and M.~P Sigalas.
\newblock The pseudo-$\pi$ model of the induced magnetic field: {F}ast and
  accurate visualization of shielding and deshielding cones in planar
  conjugated hydrocarbons and spherical fullerenes.
\newblock {\em Phys. Chem. Chem. Phys.}, 21:6150--6159, 2019.

\bibitem{hajgato2006}
B.~Hajgat{\'o}, M.~S. Deleuze, and K.~Ohno.
\newblock Aromaticity of giant polycyclic aromatic hydrocarbons with hollow
  sites: {S}uper ring currents in super-rings.
\newblock {\em Chem.: Eur. J.}, 12:5757--5769, 2006.

\bibitem{sakamoto2014}
K.~Sakamoto, N.~Nishina, T.~Enoki, and J.-I. Aihara.
\newblock Aromatic character of nanographene model compounds.
\newblock {\em J. Phys. Chem. A}, 118:3014--3025, 2014.

\bibitem{Ivanistsev16}
V.~Ivani\v{s}t\v{s}ev, T.~M\'{e}ndez-Morales, R.~M. Lynden-Bell, O.~Cabeza,
  L.~J. Gallego, L.~M. Varela, and M.~V. Fedorov.
\newblock Molecular origin of high free energy barriers for alkali metal ion
  transfer through ionic liquid–graphene electrode interfaces.
\newblock {\em Phys. Chem. Chem. Phys.}, 18:1302--1310, 2016.

\bibitem{Gomez-Gonzalez18b}
V.~G\'{o}mez-Gonz\'{a}lez, B.~Docampo-\'{A}lvarez, J.~M. Otero-Mato, O.~Cabeza,
  L.~J. Gallego, and L.~M. Varela.
\newblock Molecular dynamics simulations of the structure of mixtures of protic
  ionic liquids and monovalent and divalent salts at the electrochemical
  interface.
\newblock {\em Phys. Chem. Chem. Phys.}, 20:12767--12776, 2018.

\bibitem{Merlet13c}
C.~Merlet, B.~Rotenberg, P.~A. Madden, and M.~Salanne.
\newblock Computer simulations of ionic liquids at electrochemical interfaces.
\newblock {\em Phys. Chem. Chem. Phys.}, 15:15781--15792, 2013.

\bibitem{Fedorov14}
M.~V. Fedorov and A.~A. Kornyshev.
\newblock Ionic liquids at electrified interfaces.
\newblock {\em Chem. Rev.}, 114:2978--3036, 2014.

\bibitem{Schulman97}
J.~M. Schulman and R.~L. Disch.
\newblock Thermal and magnetic properties of coronene and related molecules.
\newblock {\em J. Phys. Chem. A}, 101:9176--9179, 1997.

\end{thebibliography}

\end{document}